\documentclass[runningheads]{llncs}
\usepackage{graphicx}
\begin{document}
\title{The anatomy of a Web of Trust: the Bitcoin-OTC market}
%
%
\author{Bertazzi Ilaria\inst{1}  \and
Huet Sylvie\inst{1} \and
Deffuant Guillaume\inst{1} \and
Gargiulo Floriana\inst{2}}
\authorrunning{F. Author et al.}
%
\institute{IRSTEA, 9 Avenue Blaise Pascal, 63170 Aubière, France  \and CNRS, GEMASS, 20 rue du Berbier du Mets, 57013 Paris, France
}
\maketitle              
\begin{abstract}

\end{abstract}
Bitcoin-otc is a peer to peer (over-the-counter) marketplace for trading with bitcoin crypto-currency. To mitigate the risks of the p2p unsupervised exchanges, the establishment of a reliable reputation systems is needed: for this reason, a web of trust is implemented on the website. The availability of all the historic of the users’ interaction data makes this dataset a unique playground for studying reputation dynamics through others’ evaluations. We analyze the structure and the dynamics of this web of trust with a multilayer network approach distinguishing the rewarding and the punitive behaviors. We show that the rewarding and the punitive behavior have similar emergent topological properties (apart from the clustering coefficient being higher for the rewarding layer) and that the resultant reputation originates from the complex interaction of the more regular behaviors on the layers. We show which are the behaviors that correlate (i.e. the rewarding activity) or not (i.e. the punitive activity) with reputation. We show that the network activity presents bursty behaviors on both the layers and that the inequality reaches a steady value (higher for the rewarding layer) with the network evolution. Finally, we characterize the reputation trajectories and we identify prototypical behaviors associated to three classes of users: trustworthy, untrusted and controversial.
\section{Introduction}
Bitcoin, perhaps the most hip cripto-currency currently available, consists in a purely peer-to-peer version of electronic cash; it is used for on line payments to be sent directly from one trading party to the other without going through a financial institution \cite{nakamoto}. In such an anonymous, decentralized exchange system, mutual trust plays a crucial role in determining the feasibility of a vivid marketplace. Digital signatures and  peer-to-peer network of data sharing over transactions provide part of the solution, in which the second one is specifically thought as a tool to prevent any double spending behavior. However in over-the-count markets, where alongside with Bitcoins, other crypto and real currencies and other types of goods are exchanged peer to peer, without a central control, a user-generated system to establish the credibility of individuals, as buyer and sellers, is needed. It is not only to guarantee the actual transaction of value, but also to monitor the goodness of the global exchange, and the correct behavior of anonymous users.

In this specific case, the platform considered here, the Bitcoin-OTC market \cite{BTC},  goes beyond the already well known user-reputation mechanisms. These classical mechanisms are implemented in many system of on line platforms (ratings, gradings, reviews, etc…) that are now ubiquitous for any form of exchange, both virtual and physical, and sometimes even human interactions, from restaurant\cite{tassos} to hosting \cite{adamic} and carpooling services, and so on. In our case, a recently rising system of user reputation is implemented: the Web of Trust. \\
Web of Trust on a p2p structure is originated in cryptography, where it defined a structure of public and private key exchange, to establish the authenticity of the binding between a public key and its owner. It consists in a decentralized trust model (alternative to the centralized one which relies on a certificate authority). The idea is that, as time goes on, users will accumulate keys from other people that they may want to designate as trusted introducers. Everyone choose their own trusted introducers. Users gradually accumulate and distribute with their key a collection of certifying signatures from others. This will create a decentralized fault-tolerant web of confidence for all public keys. The scheme is flexible and leaves trust decisions in the hands of individual users \cite{car}.
Going from the realm of cryptography to the web 2.0 applications, the idea beyond the Web of Trust model is to provide a way to grant mutual trust between buyers and sellers that have no direct interactions, but it is based on the experience other users: the peculiarity lies in the fact that the "intermediary" is already in the proximity of both users.
The Web of Trust is created by the complex set of actions and interactions of the users, and the emergent structure it assumes is the resultant of the collective intelligence the virtual community can put in place. Understanding the structure and the dynamics of this kind of objects is therefore crucial both for analyzing the robustness of the OTC markets and, more in general, as a controlled playground to understand the reputation formation mechanisms. Different simulation models exist to explain the formation of reputation hierarchies in a society \cite{leviatano,precisino} and several experimental studies \cite{vdr}, online and offline have been realized in the last years. But the data-driven studies in this field remain rares \cite{epinions}, because of the difficulties to find available data on peer-to-peer interactions. While the recent literature on user-item evaluation system is enormous, due to the amount of available data in commercial and Q\&A platforms, the Bitcoin-OTC database is one of the extremely few cases allowing us to study, from the beginning to the end, the evolution of reputation in a community as the direct effect of others' evaluations.  

The specific features of the Bitcoin-OTC dataset that require careful attention when analysis is performed are mainly two: firstly, the fact that “reputation” here is not seen as direct effect of the entire interaction history of a user, but actors perceive another user's trust relying only on the ratings of people she trusts in the first place. The \emph{gettrust} command on which this mechanism is based shows the cumulative trust for a person coming from people that users trust directly, capped by how much they are trusted. The second feature is given by the website guidelines on what ratings to hand out to people: +1 is the first rate to be given to a new person with which a user had a good transaction, as the number of well gone interactions happened between the same couple of users increases, the rate may be updated upwards, where +10 is to be reserved for \emph{$\ll$close friends and associates you know in person$\gg$},  whereas negative rates are given after a not so good, or really terrible transaction. Other considerations like the size of transactions, the nature of relationship and interaction, length of history, etc., may have an impact.\\
Because of the specificities of the dataset and the relative importance of such feature also for other similar datasets, we approached them via a  multilayer network approach \cite{multiplex} in order to understand the different mechanisms in action, as we will justify in the next section.\\
The paper then continues as follows: section two presents the data structure and the analytical method used, section three presents the results in terms of static properties of the ratings network, temporal activity of the interactions over 6-years observations, and network dynamical properties. Conclusions follows with some behavioral hypothesis that may explain the presented results.

\section{Data and methods}
Bitcoin-otc is a peer to peer (over-the-counter) marketplace for trading with bitcoin cripto-currency. To mitigate the risks of the p2p unsupervised exchanges, a Web of Trust is implemented to have access to the counterpart's reputation before a transaction, as presented above. In this web of trust system a user $i$ can rate another user $j$ with an integer score $s_{ij}$ varying between -10 to 10. The dataset has been directly mined by the web page \cite{BTC}. The dataset contains 5,878 users and 35,795 ratings exchanged between 2011 and 2017. 

The website proposes a behavioral rule concerning the scores: it is suggested to assign the score 1 to a positive rating if the rater met the rated only once and to reinforce using higher scores in case of repeated meetings. This rule generates a high prevalence of $s_{ij}=1$ but, at the same time, it is frequently broken by the users' behaviors: repeated interactions among the users remain extremely rare and high scores are assigned also at the first interaction. 

\begin{figure}
\begin{center}
\includegraphics[width=0.6\textwidth]{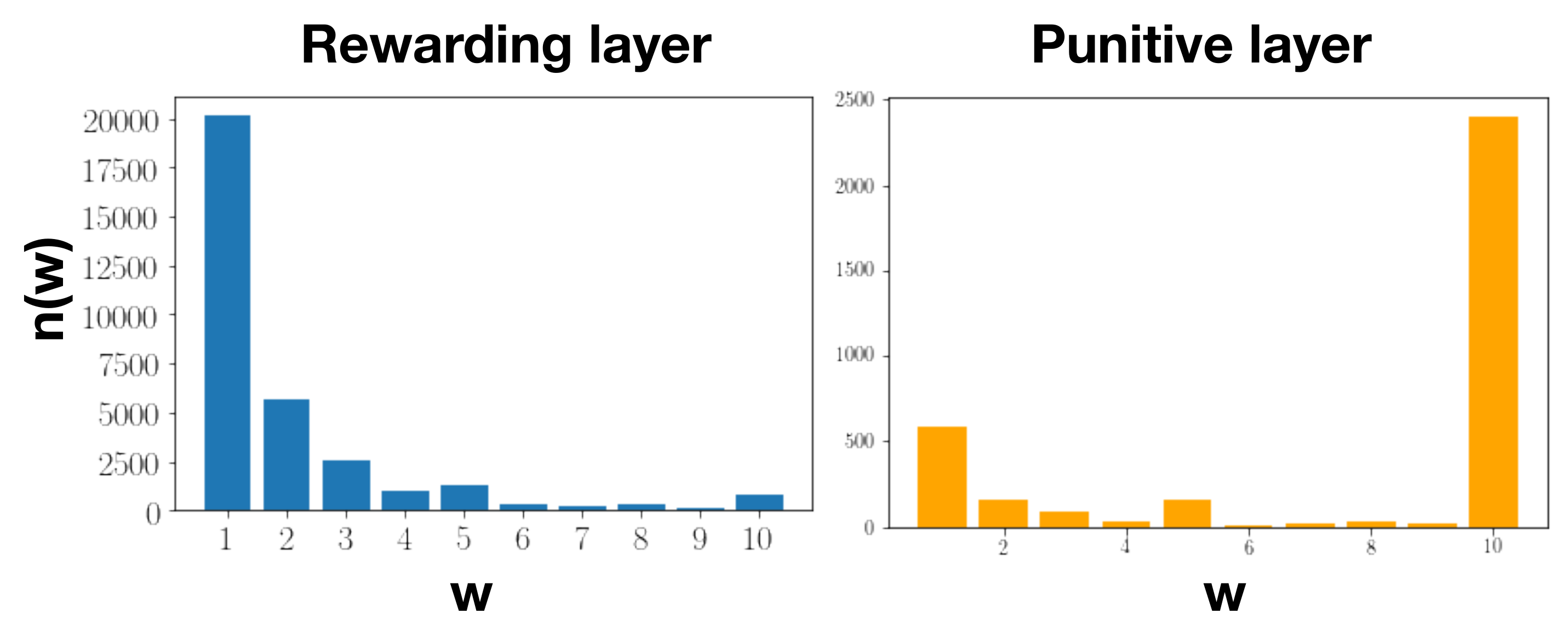}
\caption{Weight distribution for the rewarding (left plot) and the punitive (right plot) layer} \label{fig0}
\end{center}
\end{figure}

We assume that the socio-psychological micro-mechanisms governing rewarding and punitive ratings could be a priori different. 
Due to this reason we study the web of trust as a multiplex weighted directed network with two different layers: the rewarding layer, $L^+$, containing only the positive scores and the punitive layer, $L^-$, containing the negative scores. \\
Each user is therefore identified with a node both on the rewarding and the punitive layer. On both the layers the weighted edges are labelled with the time of the interaction and with the absolute value of the score associated ($w_{ij}=s_{ij}$ for the rewarding layer and $w_{ij}=-s_{ij}$ for the punitive). \\
As usual in Webs of Trust \cite{epin1,epin2}, the number of edges on the rewarding layer ($N_e^+=32305$) is much higher than in the punitive one($N_e^-=3490$). As we can observe in Fig.\ref{fig0}, also the distribution of the scores is different between the two layers: in $L^+$, due to the norm suggested by the website, the score $s=1$ is dominant, while in $L^-$ in order to emphasize the punitive gesture the score $s=-10$ is the most frequent.
To each node $i$  we associate the following of values:
\begin{itemize}
\item \bf{In-degrees=number of received scores on the two layers}
\begin{equation}
\vec{k_{in}}(i)=(k_{in}^+(i),k_{in}^-(i))
\end{equation}
\item \bf{Out-degrees=number of given scores (activity) on the two layers}

\begin{equation}
\vec{k_{out}}(i)=(k_{out}^+(i),k_{out}^-(i))
\end{equation}
\item \bf{Positive and Negative reputation}
\begin{equation}
\rho^+(i)=\sum_jw_{ji}^+,\rho^-(i)=\sum_jw_{ji}^-
\end{equation}
\end{itemize}

Finally, for each user we define the global reputation, as it is reported and visible for all the users on the website:
\begin{equation}
\rho(i)=\rho^+(i)-\rho^-(i)
\end{equation}

\section{Results}

\subsection{Static properties}
\begin{figure}
\includegraphics[width=\textwidth]{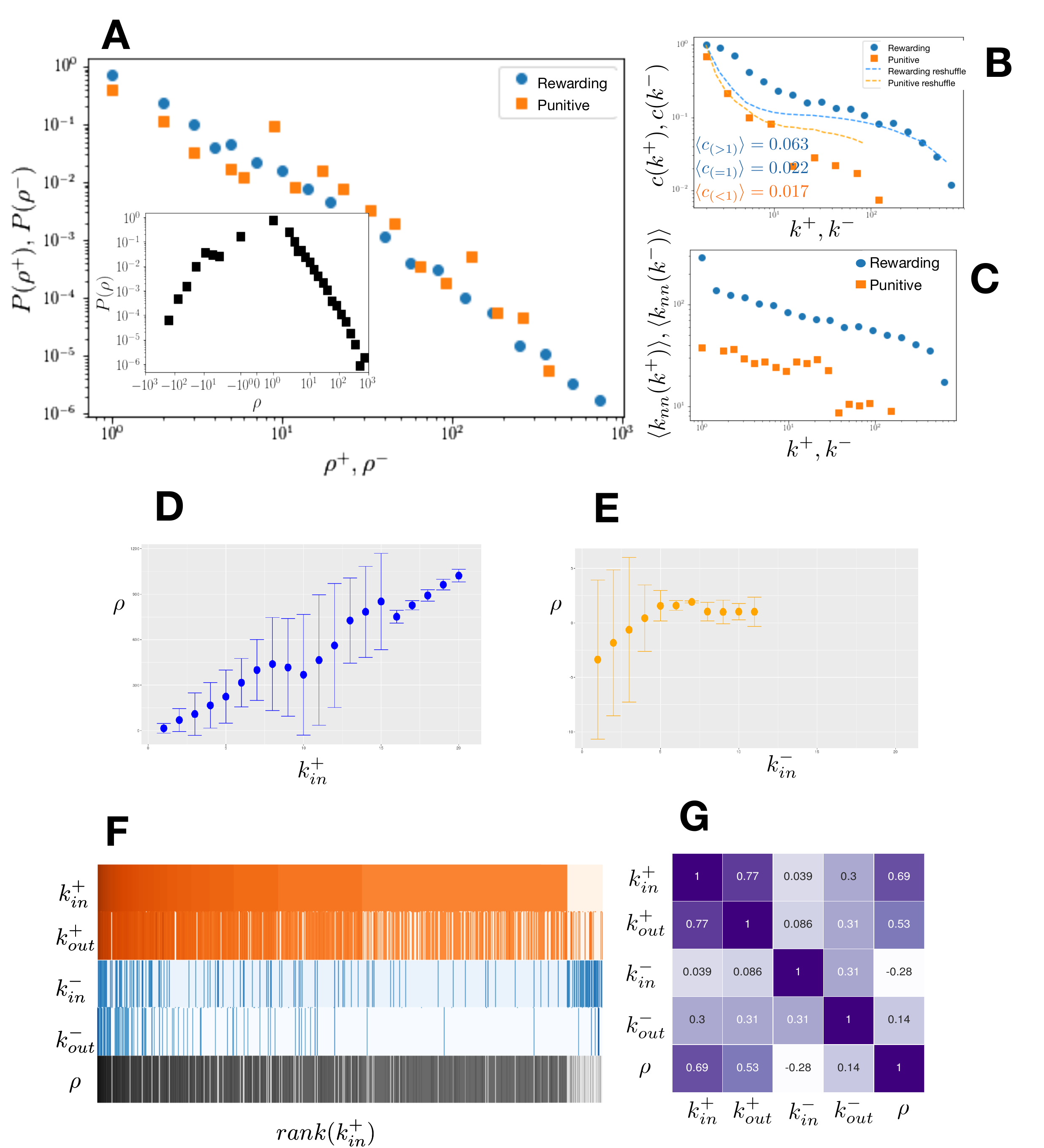}
\caption{Static properties. Plot A: distribution of the positive and the negative reputation. Inset: Global distribution of the reputation. Plot B: Clustering coefficient spectrum for the rewarding and the punitive layer, compared with the spectrum for a randomized version of the network. Plot C: Average neighbors degree spectrum for the two layers. Plot D: mean and standard deviation of the reputation as a function of the in-degrees of the rewarding layer. Plot E: mean and standard deviation of the reputation as a function of the in-degrees of the punitive layer. Plot F: In and out degrees of the nodes on the two layers and reputation, with the nodes sorted according to the in-degree ranking. Plot G: Kendall--Tau index between the node rankings for in and out degrees on the two layers and for reputation.} \label{fig1}
\end{figure}

We start analyzing the aggregate properties of the network, namely the case where the time label of the network edges is not considered. 
In Fig\ref{fig1}A we analyze the probability distributions of the positive and of the negative reputation. We notice that, notwithstanding the significant differences between the score distributions and the number of edges of the two layers, the positive and negative reputations follow the same power law distribution. In the inset, we displayed the distribution of the global reputation of the users $\rho$. The global reputation is definitely not symmetric, showing non-trivial interactions between the two layers.\\
Clustering coefficient measures the tendency to form triangles in a network structure: for a node $i$, a clustering coefficient  $c(i)=1$ signifies that in its neighborhood all the triples of nodes are connected as a triangle, $c(i)=0$, on the contrary, implies a  star topology (without triangles). The clustering coefficient of the nodes as a function of the total degree  (for the undirected  and unweighted version of the network) is displayed in Fig.\ref{fig1}B. We can observe that, for the rewarding layer, the clustering is in general higher than for the punitive layer, above all for high degree nodes. Comparing each layer with a randomized reshuffling of the network that maintains the degree distribution (configuration model), we observe that the clustering coefficient is higher than the null case for the rewarding layer and lower for the punitive. This is coherent with the balance theory suggesting that, in a triadic structure, three negative interactions, are not balanced (\cite{socialbalance}\cite{socialbalance2}). For investigating the role of the norm proposed on the website concerning the score 1, we further divided the $L^+$ in a layer with $w_{ij}>1$ and another one with $w_{ij}=1$. The average value of the clustering for the first case is $\langle c_{(>1)}\rangle=0.063$, while for the second case it is lower $\langle c_{(=1)}\rangle=0.022$. This suggests that the breaking of the norm could be due to triangular structures: we can trust someone not only because of the interaction we directly had but also because of the trust our friend address to her/him. \\ 
In Fig.\ref{fig1}C we study the average degree of the neighbors as a function of the node degree. For both the layers this indicator decreases with the degree. This indicates a typical disassortative mixing (low degree users are preferentially connected to high degree ones and vice-versa), a behavior extremelly diffused in social networks. \\
After the statistical properties of the network we concentrated on the ranking of the nodes according to the different attributes. In particular we concentrated to the values of $k_{in}^+,k_{in}^-,k_{out}^+,k_{out}^-$ and $\rho$. In Fig.\ref{fig1}F we represented these quantities according to the ranking for  $k_{in}^+$, the number of positive ratings received. The nodes with a high in-degree in the rewarding layer usually are the most active (high out-degree both on the rewarding and the punitive layer), and clearly have an high reputation. Notice however that the nodes higher in the $k_{in}^+$ ranking also receive several negative scores ($k_{in}^-$). In Fig.\ref{fig1}G we compared the ranking of the nodes according to these measures, using the Kendall-Tau index. We observe that the reputation is clearly correlated with the number of received scores in the rewarding layer ($k_{in}^+$) and with the rewarding activity ($k_{out}^+$). On the contrary the punitive activity ($k_{out}^-$), and the number of incoming negative scores ($k_{in}^-$) are scarcely correlated with all the other measures.\\
To reinforce such findings we represent the mean and standard deviation of $\rho$ of each $k_{in}$ level in Fig.\ref{fig1}D and \ref{fig1}E for both layers. In this case, it is even more clear that users' reputation is positively related to incoming interactions of the rewarding layer, whereas it is not so clearly correlated on the punishing one (it grows then becomes stable).

\subsubsection{Users' categorization}
\begin{figure}
\begin{center}
\includegraphics[width=0.9\textwidth]{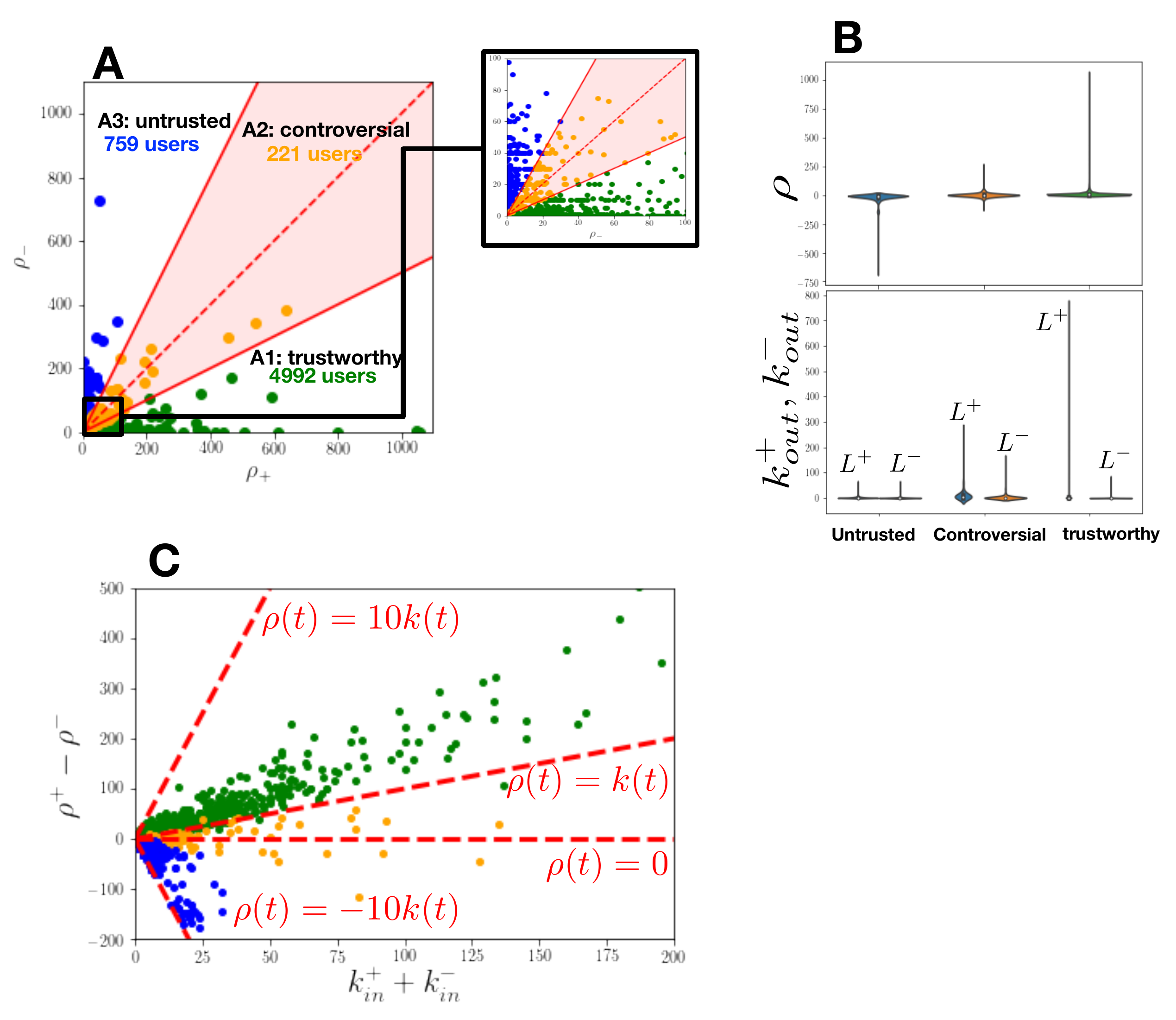}
\caption{Plot A: Identification of trustworthy (green), untrusted (blue) and controversial users (orange) with a zoom on the lower values. Plot B: Violin plot of the normed distributions of the reputation (upper plot) and of the activities (lower plot) for the three categories of users. Plot C: reputations of each user as a function of the sum of the in-degrees on the two layers. The colors of the points represent the users' categories (trustworthy, untrusted, controversial). The red lines represent limit reputation growth scenarios: $\rho(t)=10k_{in}(t)$ is the case when, at each interaction, the maximum score $s=10$ is received. $\rho(t)=k_{in}(t)$ is the case when, at each interaction, the score $s=1$ is received,  $\rho(t)=-10k_{in}(t)$ is the case when, at each interaction, the minimum score $s=-10$ is received } \label{fig2}
\end{center}
\end{figure}

We analyze now the properties of the users between the two layers, and in particular we analyze the position of the users in the space ($\rho^+,\rho^-$). We divide the plane in three areas, as described in Fig\ref{fig2}A: $A1=[\rho^+<0.25\rho^-]$, $A2=[0.25\rho^-<\rho^+<0.75\rho^-]$, $A3=[\rho^+>0.75\rho^-]$.

The users in the first area, $A1$, have an high positive reputation and a low negative, therefore, in this area we can place the \emph{trustworthy} users.
On the contrary the users in the third area, $A3$, are \emph{untrusted}, having a low positive and an high negative reputation. Finally, the users in the second area, $A2$, are \emph{controversial} having similar values for positive and negative reputation.
The largest part of the users are trustworthy. 
In the higher plot of Fig.\ref{fig2}B we can observe the distribution of the global reputations in the three different areas. Not surprisingly the untrusted users have a negative reputation and the trustworthy ones a positive one. The reputations associated to the controversial are lower. More interestingly, we can observe that the untrusted users have in general a lower activity ($\vec{k_{out}}$) on both the layers. The controversial users have a similar activity on the two layers and in general have the highest activity on the punitive layer. Finally the trustworthy users are extremely active for rewarding and much less for punishing. We can argue that the untrusted users are like "trolls" appearing and cheating one or more users just one time. In such a way they fast construct their negative reputation and after disappear. Controversial users are real users that gain and give negative scores according to more complex mechanisms that better mimic the formation of reputation in the human society. In this sense, their activity on the punitive layer could be interpreted as a reciprocation of one or more negative scores.\\
In Fig.\ref{fig2}C we finally analyze the reputation as a function of the aggregated in-degree (the sum of the number of ratings received on each layer). In an ideal case, if at each interaction the score $s=1$ is received, as it should be if the users were following the rule proposed by the website, the points should be placed on the line $\rho(t)=k_{in}(t)$. On the contrary, if the maximum or minimum scores would be always given, the points should be placed respectively on the lines $\rho(t)=10k_{in}(t)$ and $\rho(t)=-10k_{in}(t)$. We can observe that the trustworthy users are located slightly above the $\rho(t)=k_{in}(t)$ line, while the controversial ones slightly beyond.
On the contrary, the untrusted users are mostly located on the   $\rho(t)=-10k_{in}(t)$ line. We will come back to this point in the last section when analyzing the reputation trajectories. 

\subsection{Temporal activity}
\begin{figure}
\includegraphics[width=\textwidth]{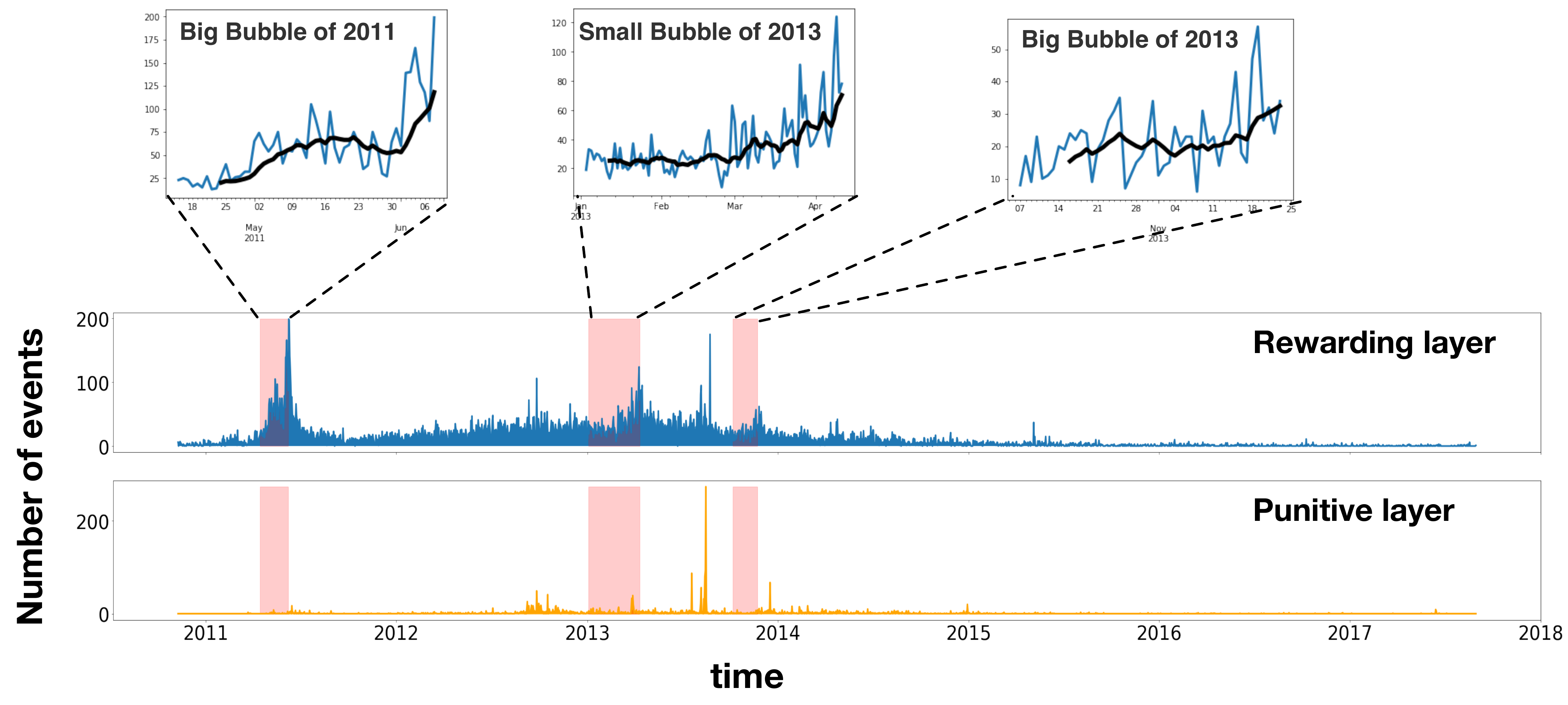}
\caption{Number of rating events time series for the rewarding (in blue) and the punitive (in orange) layer. The shadowed areas represent the three major bitcoin bubbles and the upper subplots the zoom of the time series behavior in the bubble periods.} \label{fig3}
\end{figure}

In this section we analyze the temporal patterns associated to the network activity. Aggregating the events on a daily time window, in Fig.\ref{fig3} we show the number of edges, namely rating events, in time. The rewarding layer shows some evident activity peaks. The shadowed areas in the figure represent the three major bitcoin bubbles and we can observe that the number of events in the web of trust, reasonably corresponding to an increase of the trading activity on the web site, is a growing function in the bubble periods. The punitive layer, on the contrary has in general a very low activity, but important activity burst localized on small time intervals, not related to the bubbles, appear. In particular an extremely large punitive activity can be identified between the small bubble of 2013 and the big bubble of 2013. 

\begin{figure}
\includegraphics[width=\textwidth]{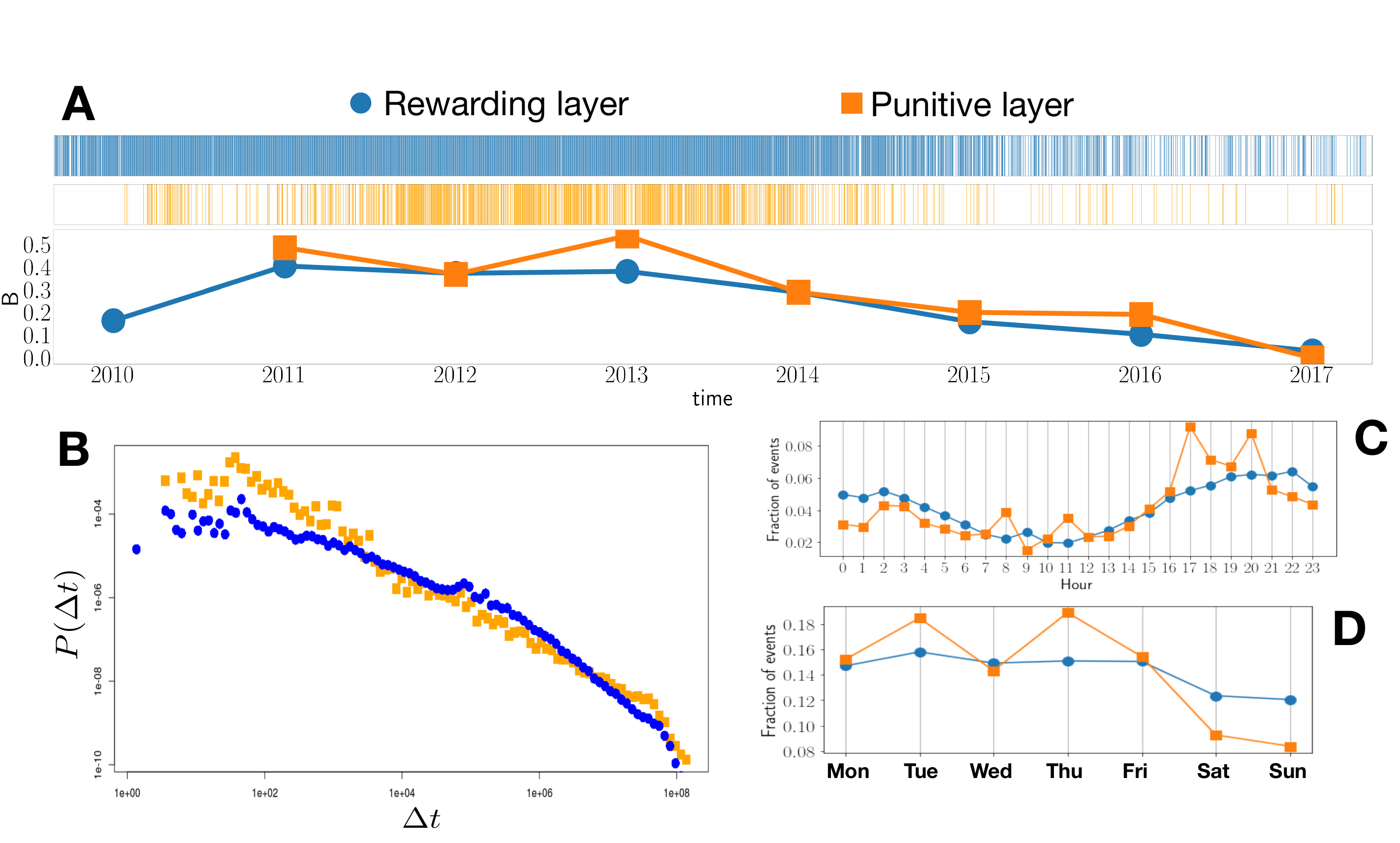}
\caption{Plot A: Event plot for the two layers and burstiness coefficient for all the years in the data collection. Plot B: interevent distribution. Plot C: fraction of events as a function of the hour of the day (Greenwich time). Plot D: Fraction of events as a function of the day of the week.} \label{fig4}
\end{figure}

In the upper plots of Fig.\ref{fig4}A we show the event plot of the two layers: a colored trait represents the days where some activity has been present. The rewarding layer has been active almost continuously between 2012 and 2015. The event plots suggest the presence of burstiness in the system activity: a typical behavior observed in different complex systems, implying activity peaks concentrated on short periods alternated to long inactivity periods \cite{barabasi}.

In Fig.\ref{fig4}B we study the inter-event distribution, where the inter-event time $\Delta t$ is defined as the time elapsed between two subsequent rating events reaching a certain user. The power law behavior of the inter-events is a fingerprint of burstiness, contrarily to the exponential distribution that would be originated by a Poisson process. The measure in the lower plot of Fig.\ref{fig4}A is an indicator of bursty behavior, defined in \cite{barabasi}: B=1 corresponds to a bursty behavior, B = 0 is neutral, and B = −1 to a regular behavior. We analyze this indicator, separately, for all the years of the data collection. Confirming the observation from the power law inter-events distribution, a bursty behavior is present, above all during the period of major activity of the website.\\
The behavior on the two layers is similar if we look at the inter-event times. However, from the measure of $B$, we can see the bursty behavior is slightly more pronounced for the punitive layer. \\
The rating activity follows circadian patterns (Fig.\ref{fig4}C): since most of the Bitcoin-OTC users are located in the US (GTM-6), and that the times in the data are relative to Greenwich time, we can observe that the activity is more intense during the day than during the night. 
In particular punitive behaviors are more frequent around lunch time, between 11am and 2pm. \\
Also week cycles are present (Fig.\ref{fig4}D),  characterized by a lower activity in the week-end. In particular negative behaviors are extremely rare during the week-end. 

\subsection{Network dynamical properties}
\begin{figure}
\includegraphics[width=\textwidth]{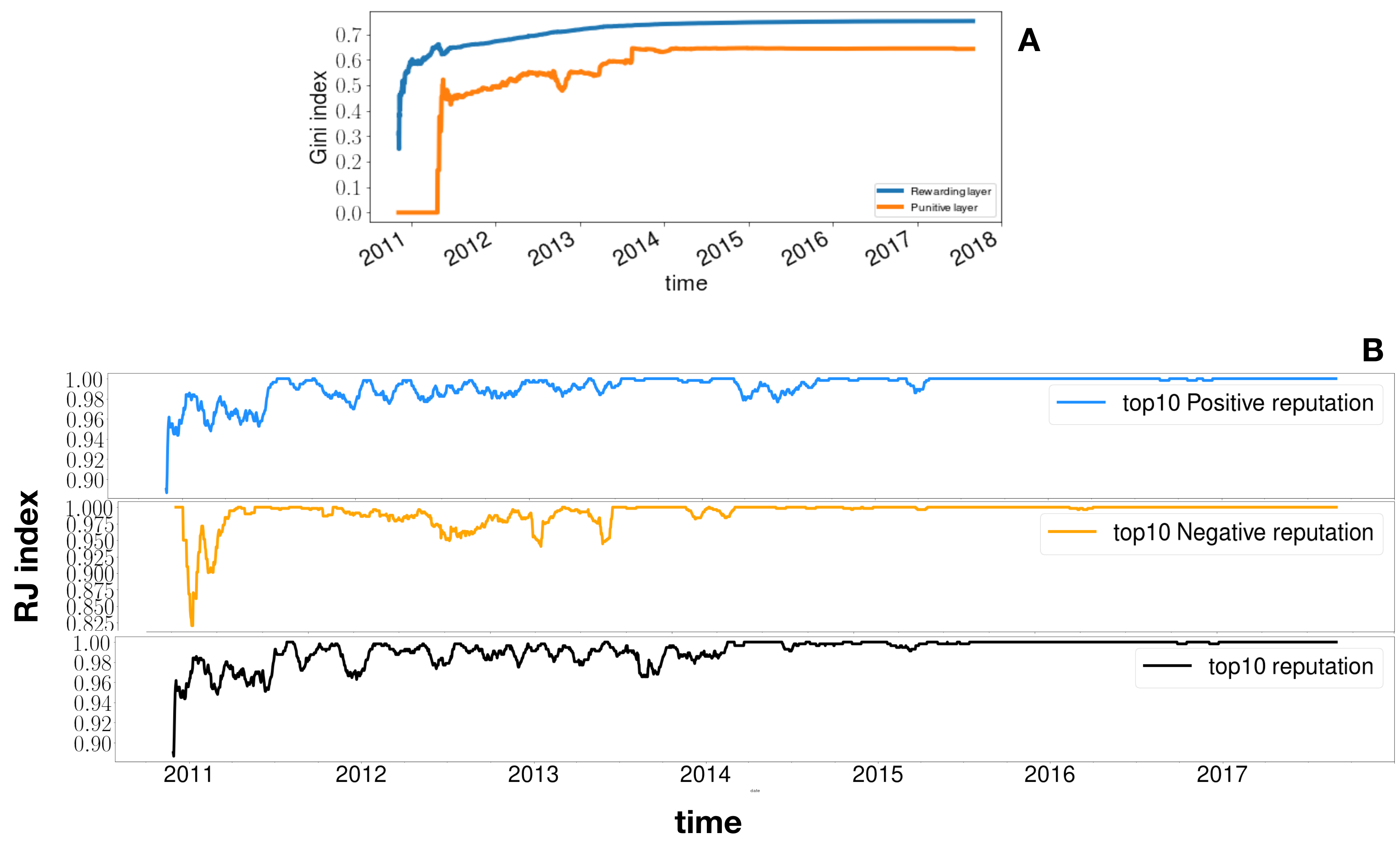}
\caption{Plot A: Gini index for positive and negative reputations. Plot B: Extended Jaccard index for the comparison between the top 10 lists at time $t$ and time $t+1$ for positive, negative and global reputation.} \label{fig5}
\end{figure}

We first analyze some global measures to understand how positive and negative reputation is distributed in the network's layers. First we measure the Gini index for positive and negative reputation. Gini index is an indicator of inequality and it ranges between $Gini=0$, when an uniform distribution is present, to $Gini=1$ when all the analyzed quantity is owned by a single individual. Measuring the Gini index for each day of the evolution (Fig.\ref{fig5}A), we observe that both for positive and negative reputations, it increases fast and after it reaches a stable value. The inequality, at the stable point, is in general quite high and larger for the rewarding layer ($G^+_{eq}=0.75$,$G^-_{eq}=0.6$). \\
Secondly we focus on the reputation ranking, for positive, negative and global reputation. In Fig.\ref{fig5}B we compare the top 10 of the rankings at time $t$, $top10_t$, with the top 10 of the ranking at time $t+1$, $top10_{t+1}$. For comparing the top 10 of the rankings we use the extended Jaccard index, introduced in \cite{matematici}. As for the original Jaccard index the modified version is such that J(rank1,rank2)=1 when the rankings rank1 and rank2 are completely equivalent, and gives a value 0 when these latter are not correlated at all.  After a transient period the top 10 lists become quite stable, even if showing continuous fluctuations. Surprisingly the fluctuations of the global reputation do not follow the fluctuation of the positive reputation, meaning the interplay with the punitive layer plays a central role in the reputation dynamics. Moreover reputation top 10 shows a certain regular periodicity, for the period of maximum use of the Bitcoin-OTC site (2012-2015). \\

\begin{figure}
\includegraphics[width=\textwidth]{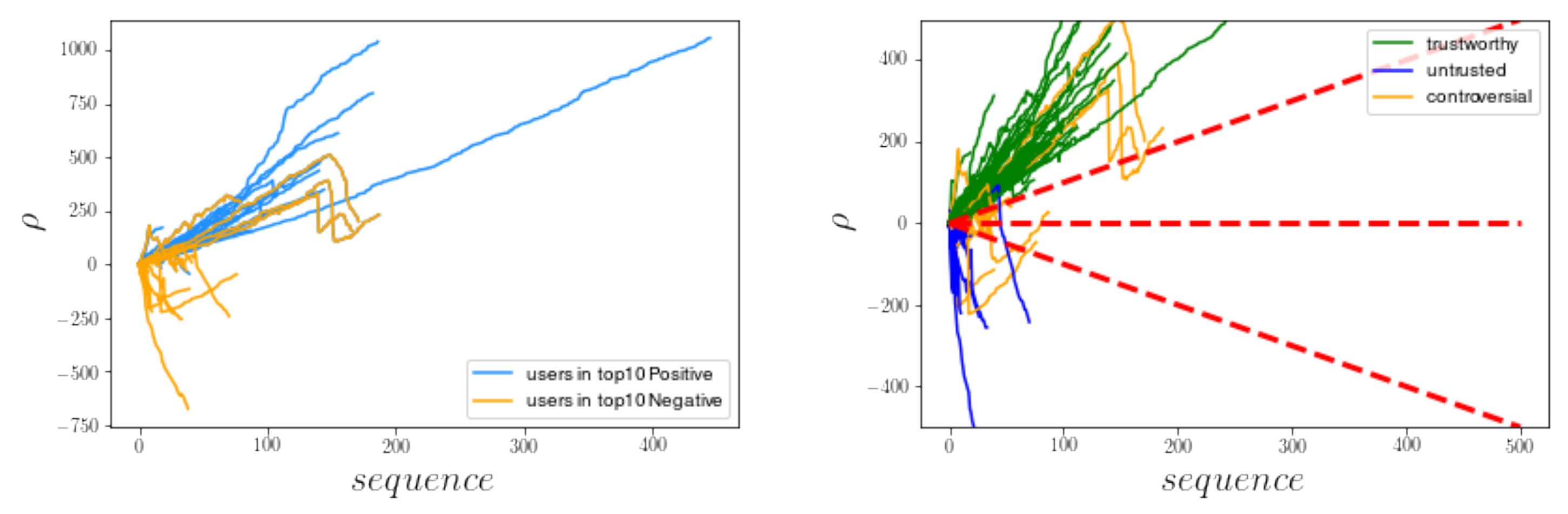}
\caption{Left plot: Flattened reputation trajectories for users in the top 10 list of positive and negative reputations. Right plot: Flattened reputation trajectories for trustworthy, untrusted and controversial users.} \label{fig6}
\end{figure}

Finally we focus on individual trajectories. For doing this we analyze, for each user, the time flattened trajectories, namely, the sequence of reputation changes (dropping the times when reputation does not change). In the left plot of Fig.\ref{fig6} we show the trajectories for the users that, at a certain point of the system evolution, entered the top 10 of positive or negative reputations. Notice that while the users in the top 10 of positive reputation have a linear growth, with a slope usually higher than the linear one, users in the top 10 of negative reputations, can in general have high global reputations since they alternate phases of growing positive reputation and phases of growing negative reputation. In the right plot of Fig.\ref{fig6} we show the trajectories for trustworthy, untrusted and controversial users, as defined previously. Untrusted users have fast linear reputation decreasing trajectories while trustworthy users have slow linear increasing trajectories. Controversial users, experience phases of reputation growth following the slow linear trend of trusted users alternated to fast reputation crashes, following the fast linear trend of untrusted users.

\section{Conclusions}
This paper aims to analyze an interesting database of an online p2p exchange community, over the mutual evaluation of users in respect to trading behaviors. We choose a Network analysis approach, with a multilayer representation, because it allows us to distinguish the specific features of rewarding and punitive behaviors.  \\
The dataset is characterized by a great majority of positive rates in respect to negative ones, where in most of the cases interactions happens only once. This leads, given the suggested rule in attributing the rates presented in the second paragraph, to a great abundance of rates = 1, but the clustering coefficient for ratings higher than 1, presented in paragraph 3.1 also underlines another aspect: the Web of Trust idea reinforces rewards even in the first encounter, when a mutual trustee is present, despite of the suggested guidelines. Looking at the ranking on in/out degree on the two layers, we can make the hypothesis that there exists a virtuous circle of activity on the rewarding layer (having a series of good rate is correlated to giving good grades), but the fact of having several negative incoming votes is not uniquely coming from retaliation behavior, as underlined by the fact that people with high $k_{in}^+$ may also have many $k_{in}^-$. In terms of accumulation of positive and negative ratings as interactions increase, we can see that, from Fig.\ref{fig1}, that where the rewarding layer behaves intuitively, the punishing one is slightly more complex to explain: this is possible to make the hypothesis that there is a strategic behaviour on the side of some users, that increases their reputation on the negative side of the network, but also have to increase the positive reputation in order to remain active in the system and find possible new interactions. \\
If we look at the specific users' trajectories, we can subdivide the vote received (which is a proxy of the goodness/viciousness of a user's trading behavior) in three categories: \emph{trustworthy, untrusted} and \emph{controversial}. Where the growth of the first group is steady and linear, the second group exhibits a non symmetric, steeper degrowth, leading to the idea that punishing rates are more firmly and quickly given than trust is granted. \emph{Controversial} behaviour on the other hand displays a mix of both trajectories.\\
Dynamically, it is to expect that the reputation distribution gets more and more unequal as time goes on because of the persistence of some users, that are active since the beginning of the time series, and the newcomers who still have to accumulate credibility. This is reinforced also if we look at the top-10 ranking over time, which becomes more and more stable.

Going beyond the analysis of the Bitcoin-OTC market, several other noticeable results, concerning human reputation studies, can be pointed out from our analysis and deserve discussions.
Firstly, our results show that reputation is gained slowly while it is lost sharply. This result has been already proven in psychological experiments like \cite{Yaniv2000} \cite{Bonaccio2006}; in particular, this studies prove that good reputation is more easily lost than gained, and authors have argued about some complementary explanations for this phenomenon, like the idea of asymmetry of trust \cite{Slovic1993}, or the heuristics used for an impression judgment: judgments are inordinately influenced by an actor's more negative attributes. This is probably related to the fact a negative information is perceived as more diagnostic of an actor's true character than positive information is \cite{Skoronski1989}.

Secondly, we have noticed that despite the rule given by the platform to rate 1 when satisfied at the first transaction, some users, relying on intermediary users, rate higher than 1 at their first transaction. This is a real peculiarity of social systems, which can't be easily governed by an a priori institution: social norms emerge from the interaction between users and are hardly predictable. This is particularly true when the behaviour to adopt is advised by trusted users as shown by the Theory of Reasoned Action \cite{Ajzen2001}. This theory has pointed out how important is the subjective norm, the norm recommended by the important others, in the decision-making regarding the behaviour to adopt. 

A final comment about \emph{controversial} users: it is possible to argue that they maintain themselves through a particular strategy. But this would imply that they have the capacity to control the system and the rate they receive, which is not really in accordance with our second noticeable result. Differently we may argue that \emph{controversial} users emerge from different strategies over the next rating to give. The anchoring bias, exhibited by \cite{Tversky1974}, is known as impacting the value given by a user, based on the value proposed by the adviser, i.e. the anchor. Then, when users base their choice of an adviser on his global reputation which is then their anchor, this can be very different from when they use their trusted users's trust as an anchor. This particular anchor can vary from the global reputation, but also from a user to another since the set of trusted users is not the same. When users have the two types of information, global reputation, and trust from their trusted users, we know that the global reputation, that we can assimilate to the descriptive norm, is not very strong in the determination of the behaviour \cite{Rivis2003}. When a user receives a negative rating, event becoming more and more probable due to the fact it has to satisfy the increasing expectations \cite{Yau-Fai1976}, she/he can enter in a negative loop, and perhaps being saved later by few people. Overall, despite the fact our work show the existence of these controversial users, we don't know much about the conditions of their emergence and maintenance. This is certainly an issue for next studies.

%
%
%
 \bibliographystyle{splncs04}
%

\end{document}